# Fluorographene: a Two Dimensional Counterpart of Teflon

R. R. Nair[1,2], W. C. Ren[1,3], R. Jalil[1], I. Riaz[1], V. G. Kravets[1], L. Britnell[1], P. Blake[2], F. Schedin[2], A. S. Mayorov[1], S. Yuan[4], M. I. Katsnelson[4], H. M. Cheng[3], W. Strupinski[5], L. G. Bulusheva[6], A. V. Okotrub[6], I. V. Grigorieva[1], A. N. Grigorenko[1], K. S. Novoselov[1,2], A. K. Geim[1,2]

[1]School of Physics & Astronomy and [2]Manchester Centre for Mesoscience & Nanotechnology, University of Manchester, Manchester M13 9PL, UK
[3]Shenyang National Laboratory for Materials Science, Institute of Metal Research, Chinese Academy of Sciences, 72 Wenhua Road, Shenyang 110016, P. R. China
[4]Radboud University of Nijmegen, Institute for Molecules and Materials, 6525 AD Nijmegen, The Netherlands
[5]Institute of Electronic Materials Technology, Wólczyńska 133, Warszawa 01-919, Poland
[6]Nikolaev Institute of Inorganic Chemistry, SB RAS, 630060 Novosibirsk, Russia

**We report a stoichiometric derivative of graphene with a fluorine atom attached to each carbon. Raman, optical, structural, micromechanical and transport studies show that the material is qualitatively different from the known graphene-based nonstoichiometric derivatives. Fluorographene is a high-quality insulator (resistivity $>10^{12}\Omega$) with an optical gap of 3 eV. It inherits the mechanical strength of graphene, exhibiting Young's modulus of 100 N/m and sustaining strains of 15%. Fluorographene is inert and stable up to 400°C even in air, similar to Teflon.**

## 1. Introduction

Extraordinary properties of graphene continue to attract intense interest that has expanded into research areas beyond the initial studies of graphene's electron transport properties.[1] One of the research directions that have emerged recently is based on the notion of graphene being a giant macromolecule that as any other molecule can be modified in chemical reactions.[2] Graphene's surface has been decorated with various atoms and molecules[3-7] but stoichiometric derivatives have proven difficult to achieve. There are two known derivatives of graphene: namely, graphene oxide[4] (GO) and graphane.[8] GO is essentially a graphene sheet densely but randomly decorated with hydroxyl and epoxy groups and obtained by exposure of graphite to liquid oxidizing agents. On a microscopic level, GO appears inhomogeneous with a mixture of regions that are pristine and densely decorated.[7] Graphane is a theoretically predicted stoichiometric derivative of graphene with a hydrogen atom attached to each carbon.[8] Graphene membranes with both surfaces exposed to atomic H exhibited a compressed crystal lattice which has served as a proof that this stoichiometric material is realizable.[9] Graphene with only one side exposed to H has a non-stoichiometric composition and, similar to graphene, exhibits metallic conductivity at room temperature ($T$). Importantly, graphene hydrogenated from either one or both sides rapidly lost H at moderate $T$,[9] which casts doubts that graphane could be used in applications where stability is required.

One way to create more stable graphene derivatives is to try using agents that bind to carbon stronger than hydrogen. Fluorine is one of such candidates and, by analogy with fluorocarbon, we refer to fully fluorinated graphene as fluorographene (FG). FG is a two-dimensional (2D) analogue of Teflon that is a fully fluorinated (FF) 1D carbon chain. Alternatively, one can consider FG as a 2D counterpart of graphite fluoride (GrF), a 3D compound used in batteries and as a lubricant.[10] Recently, mechanical cleavage[11] was attempted to extract individual atomic planes (that is, FG) from commercially available GrF but it proved surprisingly difficult and only multilayered samples were reported.[12-14] Moreover, the studied multilayers exhibited electronic and Raman properties that, as shown below, resemble partially reduced FG and are qualitatively different from the stoichiometric material reported here. The latter is a wide-gap insulator with a mechanical strength and elasticity similar to graphene and thermal stability and chemical inertness matching those of Teflon. The two-dimensional insulator complements metallic properties offered by its parent material and can be used as atomically thin tunnel barrier.

## 2. Results and Discussion
### 2.1. From Graphene to Fluorographene

We have employed two complementary approaches for obtaining FG. One is the mechanical cleavage of GrF, similar to reports[12-14]. Its monolayers were found to be extremely fragile and prone to rupture, due to many structural defects resulting from the harsh fluorination conditions used to obtain bulk GrF.[10,15] Nevertheless, we have succeeded in extracting GrF monolayers of ~1 μm in size (see Supporting Information #1) and used them in Raman studies. To prepare large FG samples suitable for most of our experiments, we have found it both necessary and convenient to employ an alternative approach in which graphene was exposed to atomic F formed by decomposition of xenon difluoride ($XeF_2$)[16] (note that, at room $T$, graphene is stable in molecular $F_2$).[10] This approach has a clear advantage with respect to possible fluorination in plasma (as employed for hydrogenation of graphene)[9] because the use of $XeF_2$ avoids a potential damage due to ion bombardment. Furthermore, fluorination by $XeF_2$ is a simple low-hazard procedure that can be implemented in any laboratory.

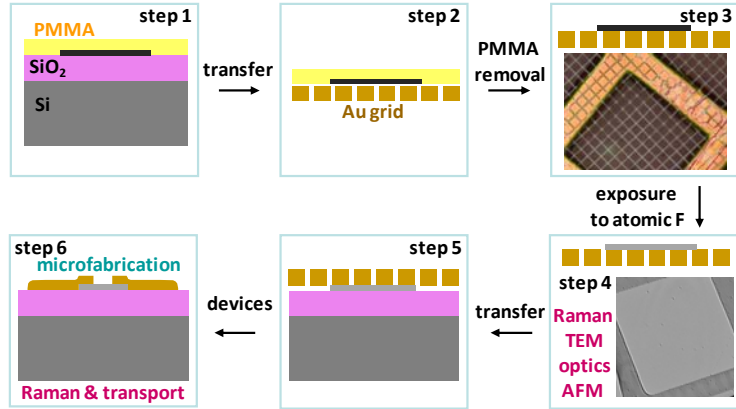

*Figure 1.* Various steps involved in our investigation. They are discussed in detail in the Experimental Section.

The processing chart to obtain FG samples used in our experiments is shown in Figure 1. In brief, we prepared large graphene crystals (>100 μm in size) by using the standard cleavage technique.[11] Because $XeF_2$ rapidly etches Si and easily diffuses through even a thick layer of amorphous $SiO_2$, it appeared impossible to use Si wafers in the fluorination procedures. Keeping in mind the necessity of using a chemically inert support and the fact that complete fluorination requires the exposure of graphene from both sides, we transferred the cleaved crystals onto Au grids used for transmission electron microscopy (TEM). To provide sufficient support for graphene, we used Au grids covered with Quantifoil, a lithographically patterned polymer film (see Figure 1). The samples were then placed in a Teflon container with $XeF_2$ and heated to 70 °C (the elevated $T$ speeded up the reaction; the use of even higher $T$ destroyed the Au grids). The resulting samples were then used for Raman, TEM and optical studies and probed by atomic force microscopy (AFM). For electrical characterization, FG was transferred from TEM grids back onto an oxidized Si wafer. The latter was done by pressing the grids against the wafer or by using the capillary transfer method.[17]

### 2.2. Raman Spectroscopy of Fluorinated Graphene

The evolution of graphene's Raman spectra due to its consecutive exposures to atomic F is shown in Figure 2. One can see that first a prominent D peak emerges. This indicates the appearance of atomic scale defects.[18,19] As the fluorination time increases, the double-resonance band (usually referred to as 2D or G' peak) disappears whereas D and G peak intensities remain approximately the same (Figure 2b). With increasing the fluorination time (a few days), all the Raman features gradually disappear. This behavior is radically different from the one observed for hydrogenated graphene, in which case the 2D band always remains strong.[9] Partially fluorinated graphene (10 to 20 h) exhibits the Raman spectra that resemble those of GO that also has comparable intensities of the G and D peaks and a relatively small 2D band.[7,20] The disappearance of all the characteristic peaks clearly proves more dramatic changes induced by fluorination in comparison with those reported for hydrogenated graphene and GO. We explain this by complete optical transparency of FF graphene to our green laser light. Indeed, according to theory,[21,22] GrF should have $E_g \approx 3.5$ eV (the gap has not been measured previously,[10] probably because the material usually comes in the form of an opaque white powder).

It is instructive to compare the observed spectrum of FG with Raman spectra of bulk GrF and a monolayer extracted from the latter (Figure 2c). One can see that, within the noise level, the former two are identical and correspond to the spectrum of partially fluorinated graphene (close to the state achieved after 20-30 h in Figure 2a). This is surprising because GrF normally exhibits fluorine-to-carbon ratios larger than unity (in our case, the ratio is $\approx 1.1$; see Supporting Information) and is assumed to be fully fluorinated.[10] The non-stoichiometric F/C ratios are due to the presence of numerous structural defects that allow more C bonds to be terminated with fluorine ($CF_2$ and $CF_3$ bonding). Our Raman data show that, despite F/C >1, GrF planes remain not fully fluorinated, and one should not use GrF spectra as a reference to achieve a FF state.

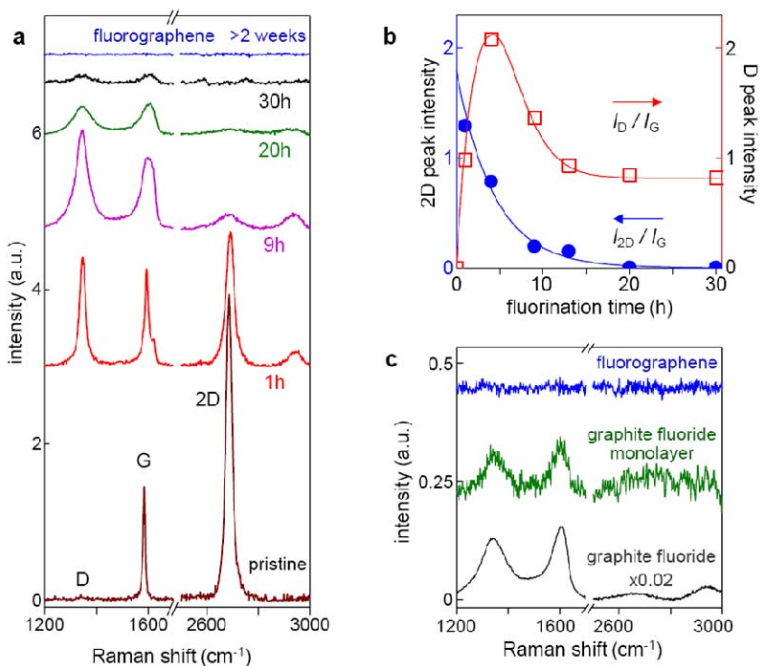

*Figure 2*. Raman signatures of FG. *(a)* – Evolution of the spectra for a graphene membrane exposed to atomic F and measured each time under the same Raman conditions. The curves are shifted for clarity. *(b)* – Intensities of the D and 2D peaks (normalized with respect to the G peak intensity) as a function of fluorination time. Solid curves are guides to the eye. *(c)* – Comparison of our FF membranes with GrF and its monolayer. The measurements were done under the same conditions. The curves are shifted for clarity, and the one for bulk GrF is scaled down by a factor of 50. For strongly fluorinated samples, a smooth background due to luminescence was removed.

**2.3. Structure and Stability**
Structural information about FG was obtained by TEM. Figure 3a shows an electron diffraction micrograph for a FF membrane. The image yields a perfect hexagonal symmetry and is similar in quality to those observed for pristine graphene.[23] The unit cell of FG is slightly expanded with respect to graphene's cell, in contrast to the case of hydrogenated graphene that showed a compressed lattice.[9] FG's lattice was found the same for all the studied FF membranes and its expansion was isotropic (no axial strain was observed[9]). Figure 3b shows histograms for the lattice constant *d* in graphene and FG. The spread in the recorded values is due to a limited accuracy of TEM in precision measurements of *d*. Nonetheless, one can clearly see that FG has a unit cell approximately 1% larger than graphene, that is, $d \approx 2.48$ Å. An increase in *d* is expected because fluorination leads to *sp*3-type bonding that corresponds to a larger interatomic distance than *sp*$^2$. However, the observed increase is smaller than that in GrF where *d* were reported to be by 2.8 to 4.5% larger than in graphite.[15,24] The smaller *d* in FG is probably due to the possibility for the 2D sheet to undergo strong interatomic corrugations if out-of-plane displacements of carbon atoms are not restricted by the surrounding 3D matrix, similar to the case of graphane that is predicted[8] to have *d* close to the one we observed for FG.

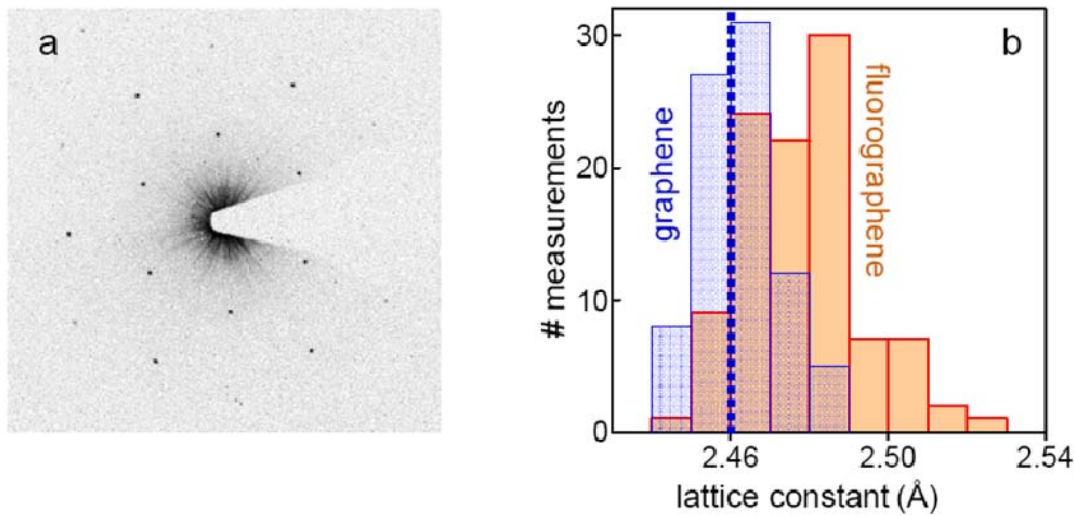

***Figure 3.*** *Transmission electron microscopy of FG. (**a**) – Diffraction pattern from a FG membrane. (**b**) – Lattice constant d measured using micrographs such as shown in (a). For comparison, similar measurements were taken for membranes before fluorination (left histogram). The dotted line indicates d for graphite.*

The Raman signatures for complete and partial fluorination in Figure 2a allowed us to study the FG stability at elevated *T* and with respect to exposure to various chemicals (also, see Supporting Information #3). For graphene fluorinated only for a few hours, the process was found to be largely reversible, so that a short annealing at 250 °C in an argon-hydrogen mixture (10% $H_2$) could restore membranes to their nearly pristine state with only a little D peak left. After more extensive fluorination (>20h), the annealing even at 450 °C could not restore the 2D peak but the D and G peaks notably grew and became similar in intensity to those on, for example, the 9h curve in Figure 2a, which indicates that a significant amount of F remained attached to the carbon scaffold. For FF graphene, its Raman spectra did not change for *T* up to 200 °C and losses of F became discernable only for prolonged annealing above 400 °C.

FG was also found to be stable in such liquids as water, acetone, propanol, etc. and under ambient conditions. The chemical stability is similar to that of graphite fluoride and Teflon, although our tests were not exhaustive. Note that we have also investigated digraphite fluoride ($C_2F$), a stage II intercalation graphite compound.[25] This material allowed relatively easy exfoliation but was unstable in any of the above liquids. Its single- and few-layer crystals were unstable even under ambient conditions, reducing rapidly to the state similar to strongly damaged graphene or reduced GO. Further information about FG's stability was obtained in transport experiments discussed below.

## 2.4. Optical Properties
The absence of Raman signatures for FG has indicated its optical transparency. Figure 4 extends this qualitative observation by showing absorption spectra of pristine, partially and FF graphene. The measurements were done for graphene deposited onto quartz wafers and then fluorinated in $XeF_2$ at 70 °C, which did not damage quartz in a moisture-free atmosphere. This method allowed us to obtain large crystals (>100 μm in size) suitable for standard optical spectroscopy. The crystals' transparency was measured with respect to the wafer.

The upper curve in Figure 4 is for pristine graphene. For light energies $E$ <2.5 eV, it exhibits a flat absorption spectrum $abs(E)$ with a "universal opacity" of $\pi\alpha \approx 2.3\%$ where $\alpha$ is the fine structure constant.[26,27] Strong deviations from this universality take place in blue, and graphene's opacity triples in peak at 4.6 eV. This is due to the fact that graphene's spectrum is no longer linear at energies close to the hopping energy of $\approx$2.5 eV and exhibits a pronounced van Hove singularity.[28-30] Note that the peak is clearly asymmetric with a low-$E$ tail, which is attributed to excitonic effects.[29,30]

For graphene fluorinated on quartz, its state was first assessed by Raman spectroscopy. Although F should be able to diffuse between graphene and quartz, [10] the concentration of atomic F underneath the graphene sheet is probably limited by its recombination into less reactive $F_2$. Accordingly, it required several days to reach the fluorination state similar to that achieved after 9 hours for membranes in Figure 2a. The partially fluorinated graphene exhibited enhanced transparency with respect to graphene over the whole E range (Figure 4) and, for visible light, its opacity fell down to $\approx$0.5%. Because impurity scattering is not expected to result in any significant decrease in optical conductivity, the enhanced transparency of the partially fluorinated state can only be explained by a gap that opens in graphene's electronic spectrum (Supporting Information #7). The remnant absorption can be attributed to microscopic regions that remain non-fluorinated, similar to the case of GO.[7]

After several weeks of fluorination, we achieved the Raman state that corresponded to the 30h curve in Figure 2a. This highly fluorinated state was found to be transparent at visible frequencies and started absorbing light only in blue (Figure 4). This proves that FG is a wide gap semiconductor with $E_g \geq$ 3.0 eV. To confirm this result, we also used the technique described in ref. [26], which analyzed images obtained in an optical microscope by using a set of narrow-pass filters. The latter approach limited our measurements to the visible spectrum but allowed the use of FF membranes directly on a TEM grid. No opacity was detected for these samples at all frequencies accessible by the microscopy approach (large symbols in Figure 4).

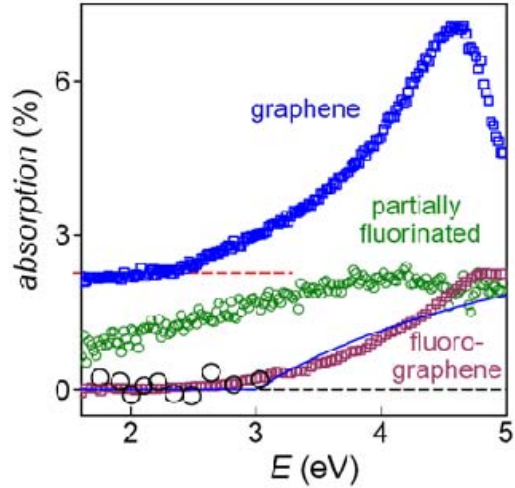

*Figure 4. Changes in optical transparency of graphene due to fluorination. The upper curve is for graphene and, within experimental error, follows the low-E data of ref. [26]. Beyond the previously reported range (<3eV), graphene exhibits an absorption peak in ultraviolet. Partially fluorinated graphene shows higher transparency (middle curve). FG is transparent for $E \leq$ 3eV but start absorbing violet light. Large open circles are measurements for FG membranes on TEM grid by using the filter spectroscopy.[26] The dashed lines indicate zero and $\pi\alpha$ opacities. The solid curve is the absorption behavior expected for a 2D semiconductor with $E_g$ =3eV.*

Unlike bulk semiconductors, 2D materials remain partially transparent even at $E$ above the gap energy. The analysis given in ref. [30] can be extended for a 2D semiconductor with a parabolic spectrum and yields $abs(E) \approx 2\pi\alpha(1-E_g/E)$ for $E \geq E_g$. This dependence is shown by solid curve in Figure 4. For a gapped Dirac spectrum, we find $abs(E) \approx \pi\alpha(1-E_g^2/E^2)$, which fits the experimental data equally well but probably is less appropriate for the graphene spectrum with such a large gap. The measured spectra could also be influenced by excitonic effects and, therefore, provide the lower bound for the real band gap of FG. Therefore, we refer to the observed cut-off as an optical gap.

**2.5. Insulating Properties**
To assess the electrical properties of FG, we transferred our samples from Quantifoil onto an oxidized Si wafer and made multi-terminal devices such as shown in Figure 5. Even weakly fluorinated graphene (with Raman spectra similar to the 1h curve in Figure 2a) was found to be highly insulating, exhibiting room-$T$ resistivity ρ in the MOhm range, that is, three orders of magnitude higher than graphene (Supporting Information #5). This clearly distinguishes fluorination from hydrogenation, with the latter resulting in little increase in ρ at room $T$.[9] The devices made from FF graphene showed no leakage current at biases $V_{SD}$ up to 10 V (within our detection limit of ~0.1 nA due to parasitic conductivities; Supporting Information #5). Taking into account that the devices had typical width-to-length ratios of 10 to 100, this sets a lower limit on FG's ρ of >$10^{12}$ Ohm at room $T$. Such a highly insulating state is in agreement with the presence of a wide bandgap.

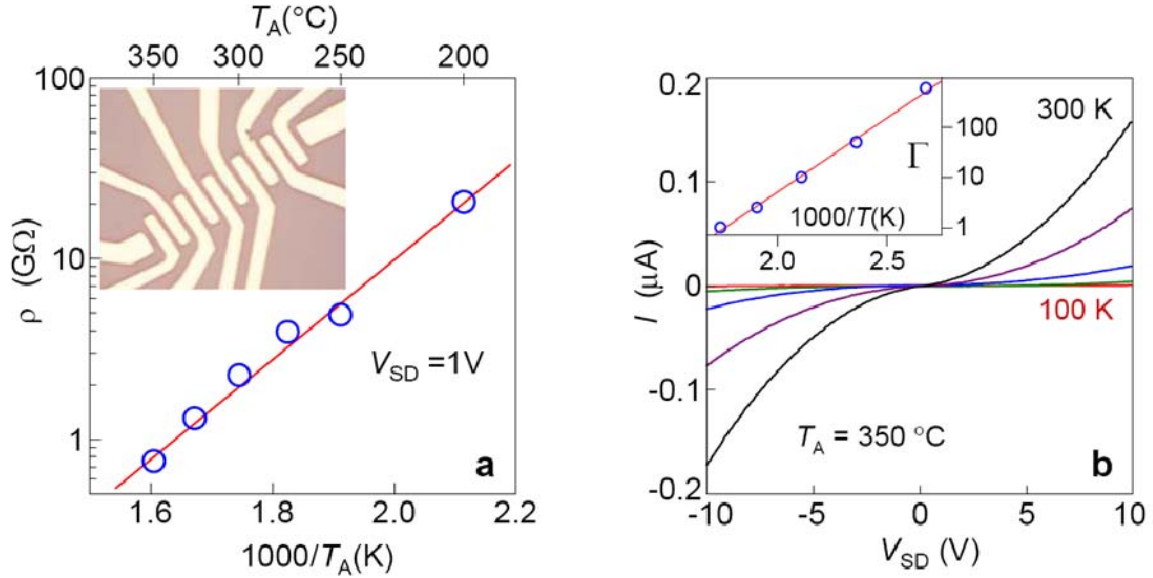

*Figure 5.* Highly stable 2D insulator. *(a) Changes in FG's ρ induced by annealing. No changes could be detected at $T_A$ below 200°C. At higher $T_A$, ρ falls below 1TΩ and becomes measurable in our experiments. Because of nonlinear I-V characteristics, the plotted ρ values were recorded for a fixed bias $V_{SD}$ of 1V (circles). For any given $T_A$, we found that it required approximately 1h to reach a saturated state. The solid line is the exponential dependence yielding $E_{des}$ ≈0.65 eV. The inset shows one of our devices with the distance between adjacent contacts of 2 μm. (b) I-V characteristics for partially fluorinated graphene obtained by reduction at 350°C. The curves from flattest to steepest were measured at T = 100, 150, 200, 250 and 300 K, respectively. The scaling factor Γ is plotted in the inset. The solid line is the best fit by exp($E_h$/T).*

Electrical measurements allowed us to study thermal stability of FG in more detail than the Raman spectroscopy. Figure 5a shows changes in electrical conductivity induced by annealing at different temperature $T_A$ in the argon-hydrogen atmosphere. No current could be detected through FG after its prolonged annealing at $T_A$ below 200 °C. At higher $T_A$, FG became weakly conductive (see Figure 5a), and at $T$ as high as 350 °C its effective resistivity ρ = $V/I$ fell down to ≈1 GΩ if we applied a large source-drain voltage $V_{SD}$ of 1 V. This behavior is in agreement with the changes observed in Raman spectra due to annealing (see above and Supporting Information #3). The ρ($T_A$) dependence in Figure 5a is well described by the functional form exp($E_{des}/T_A$) with desorption energy $E_{des}$ ≈0.65 eV. The found $E_{des}$ is notably lower than the C-F bond energy of ≈5.3eV, indicating that the initial desorption occurs from defective sites. This is consistent with the studies of GrF, which show that its decomposition is initiated at structural defects and strained regions.[31] The defect-mediated desorption is also supported by the fact that saturated states in Figure 5a are rapidly achieved after <1h of annealing and no further changes occur for longer exposures to a given $T_A$.

The electrical measurements of devices partially reduced by annealing also confirm that the material is a wide bandgap insulator, in agreement with our Raman and optical measurements. To this end, we measured I-V characteristics of FG strongly reduced at 350 °C (Figure 5b). They collapse on a single I-V curve if scaled along the $I$ axis (not shown). The found pre-factor $\Gamma$ is plotted in the inset. The $T$ dependence of $\Gamma$ is well described by the activation dependence $\exp(E_h/T)$ with $E_h \approx 0.6$ eV. The value is smaller than the minimum activation energy $E_g/2 \approx 1.5$ eV expected from our optical studies. This implies a broad band of impurity states inside the gap, which can be attributed to fluorine vacancies that appear during annealing. In this case, electron transport occurs via activation from the impurity band to the conduction or valance band, the mechanism common for semiconductors with a high density of deep dopants.[32] In the FF state (before annealing), $E_h$ should be significantly higher but we could not observe any conductivity for FG to estimate its transport gap.

We emphasize that the thermal stability of FG is higher than that of graphene, GO and even GrF. Under the same conditions, GrF starts decomposing already at 300 °C.[31] The higher stability of FG can be attributed to the absence of structural defects and little strain. As for Teflon, it undergoes slow decomposition at $T > 260$ °C and rapidly decomposes only above 400 °C.[33] Our transport measurements are sensitive to minor compositional changes (indicating none below ~200 °C) while the Raman spectra discussed above revealed notable F losses only above 400 °C (Supporting Information #3). These characteristics are very similar to those of Teflon.[33]

**2.6. Stiffness and mechanical strength**

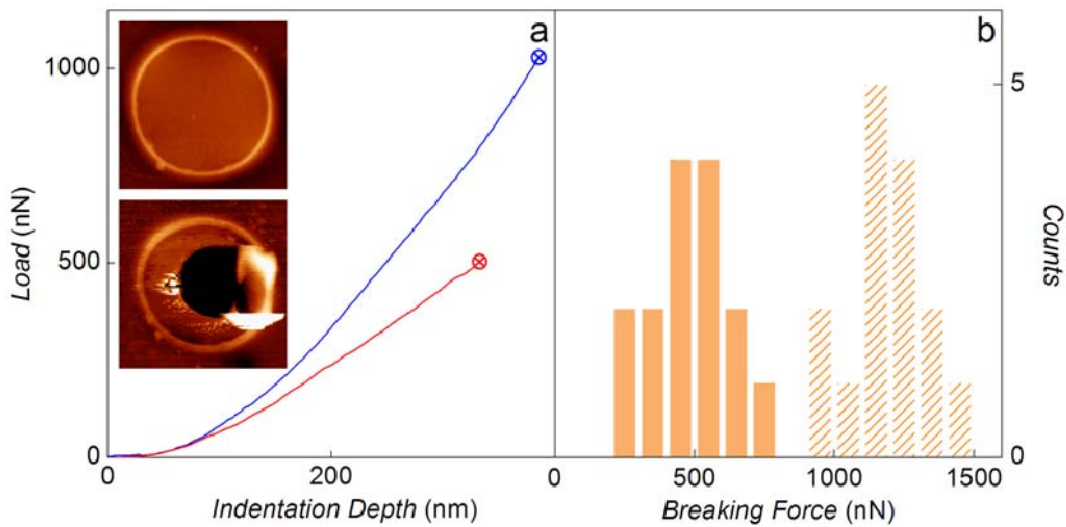

*Figure 6.* *Mechanical properties of FG. (a) – Examples of the loading curves for graphene (blue) and FG (red) membranes. Fracture loads are marked by the circled crosses. Until these breaking points, the curves were non-hysteretic. Top and bottom insets: AFM images of a FG membrane before and after its fracture, respectively. The lateral scale for the images is given by their width of 2.2 μm; Z-scale is approximately 100 nm. (b) Histogram for the breaking force for graphene (hashed) and FG (solid color). All the membranes (15 of each type) were on identical Quantifoils and punched by the same AFM tip.*

GrF contains many structural defects induced by fluorination procedures.[10,15] If our FG were similarly fragile, this would severely limit its possible applications. To gain information about the mechanical properties of FG, we have employed AFM. Quantifoil with a periodic array of circular apertures was used as a supporting scaffold (see the AFM image in Figure 6a, inset). The experimental arrangements and analysis were similar to those of Ref. [34] (Supporting Information #6). In brief, an AFM tip was positioned above the center of a FG membrane and then moved down to indent it. We recorded the bending of the AFM cantilever as a function of its displacement, and the force acting on the membrane was calculated from the cantilever's rigidity.[34,35] Figure 6a shows typical loading curves. As a reference, we used pristine graphene on identical Quantifoil grids. This allowed us to crosscheck the results and avoid systematic errors due to finite rigidity of the polymer scaffold that also responded to the load. Our analysis of the force-displacement curves has yielded Young's modulus $E$ of FG ≈100±30 N/m or 0.3 TPa[34], that is, FG is 3 times less stiff than graphene.

To measure FG's breaking strength, we indented the membranes until they collapsed (Figure 6a). The observed values for the breaking force are collected in Figure 6b. Both graphene and FG show similar histograms but graphene exhibits on average ≈2.5 times higher strength. This infers FG's intrinsic strength σ ≈15 N/m. This reduction in stiffness and breaking strength is generally expected due to the longer $sp^3$-type bonding in FG. Nonetheless, we emphasize that both $E$ and σ are extremely high in comparison with other materials (e.g., structural steel). What's more, graphene and FG can sustain the same elastic deformations σ/$E$ of ~15%. This can be readily seen from Figure 6a where both membranes broke at similar indentations. The large breaking strength of FG and the fact that it supports so high strains prove its little damage during fluorination and the practical absence of structural defects.

### 2.7. Fluorographene paper: 2D Teflon
To demonstrate that it is possible to scale up the production of FG for applications, we have fluorinated graphene laminates and graphene on SiC. Laminates were obtained by filter deposition from a graphene suspension that was prepared by sonication of graphite.[36,37] To speed up the fluorination process that involves diffusion of F between crystallites, we exposed the laminate to $XeF_2$ at 200 °C. 10 hours were sufficient to reach a saturated state that did not change with further fluorination. Note that, under the same conditions, graphite could not be fluorinated (higher $T$ are employed to produce GrF[10]). This implies that multilayer graphene present in laminates[36,37] probably remains not FF (Supporting Information #2). Graphene on SiC is discussed in Supporting Information #4.

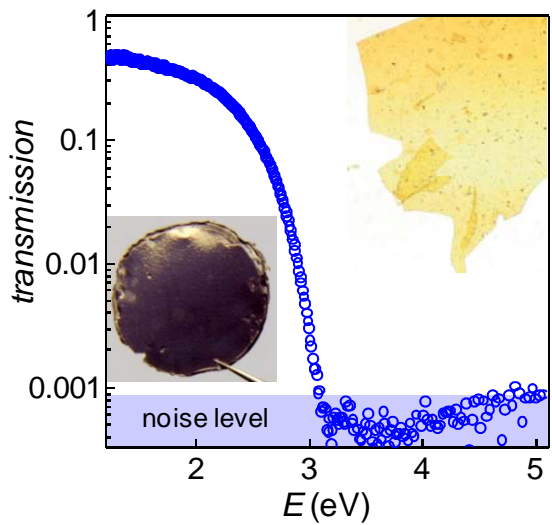

*Figure 7. Graphene paper before and after fluorination (left & right insets, respectively). The plot shows optical transparency of FG paper as a function of E for a 5 μm thick sample in the photo. The FG sample size is approximately 1 cm.*

Figure 7 shows optical photographs of a graphene laminate before and after its fluorination. The resulting material is visually distinctive from the original that is completely black with metallic shine (see the left inset of Figure 7). In contrast, FG paper is transparent and has a yellowish color that corresponds to absorption in violet (right inset). This is a direct visual proof that FG is a wide gap material. Its optical spectrum is also shown in Figure 7. The light transmission exhibits an onset at ≈3.1 eV, in agreement with the gap value inferred from the absorption spectra of individual FG crystals. The smaller gap with respect to GrF predicted to have $E_g$ ≈3.5 eV[21,22] or higher can be due to an atomically corrugated structure or excitonic effects. Note that GO has a color somewhat similar to FG but the former absorbs much more light and GO paper becomes completely non-transparent already at submicron thicknesses. Furthermore, the spectrum in Figure 7 is qualitatively different from that of GO which shows no apparent gap.[7] FG paper is strongly hydrophobic (similar to GrF) and stable under ambient conditions and at elevated $T$ as expected from our studies of individual FG crystals.

**3. Conclusions**

We have shown that the exposure of graphene to atomic F results in a stoichiometric derivative that is an excellent insulator with a high thermal and chemical stability. The optical and electrical properties of FG are radically different from those of graphene, graphene oxide and hydrogenated graphene due to a wide gap opened in the electronic spectrum. Mechanically, FG is remarkably stiff but stretchable, similar to its record-breaking parent, graphene. These characteristics rival those of Teflon and allow one to consider FG for a range of technologies, in particular those that employ Teflon rather than GrF or require better inertness and stability than unachievable for the latter compound. As for electronic applications, particularly promising seems the possibility to use FG as an atomically thin insulator or a tunnel barrier in graphene-based heterostructures such as, for example, the widely-discussed graphene double layers that have to be electrically decoupled by an atomically thin insulator. More generally, FG adds to the small family of graphene-based derivatives that previously consisted of only GO and hydrogenated graphene.

**4. Experimental Section**

*Fluorination:* Large graphene crystals were prepared on top of an oxidized silicon wafer (300 nm of $SiO_2$) by using micromechanical cleavage. Because of high reactivity of Si with atomic fluorine, we had to transfer graphene onto gold and nickel grids that could sustain the fluorination procedures. This approach also allowed us to expose graphene to F from both sides. As the first step (see Figure 1), a thin polymer layer (100 nm of PMMA) was deposited on top of the wafer with graphene crystals. The PMMA film provided a mechanical support for graphene during further processing. Then, the $SiO_2$ layer was etched away in 3% potassium hydroxide solution, which lifted off the PMMA film together with graphene crystals. After thorough cleaning in deionized water, the film floating in water was picked up onto a TEM grid (step 2 in Figure 1). Finally, PMMA was dissolved in acetone and the samples were dried using a critical point dryer. The optical micrograph in Figure 1 (step 3) shows one of our Quantifoil-Au grids. The size of the Quantifoil mesh is 7 μm, and graphene covers the whole Au cell. Graphene membranes on Quantifoil were then exposed to $XeF_2$ at 70 °C (~ 1 g in a 3 ml PTFE container). The procedure was carried out in a glove box to avoid any moisture that could result in the formation of HF. The TEM micrograph in Figure 1 (step 4) shows one of the Quantifoil cells fully covered with FG. Its presence can be witnessed as small dust particles within the aperture. We also fluorinated graphene crystals cleaved on top of quartz wafers. Such samples were used for the optical spectroscopy measurements.

*Increasing the speed of fluorination:* We have found that it is possible to significantly increase the speed by using higher $T$. To this end, a PTFE-lined stainless steel container (*Parr Instruments*) was used. The high-$T$ procedure required graphene to be placed on Ni grids that, unlike gold, could sustain $XeF_2$ at 200 °C. Using this approach, the FF state could be reached within a few hours rather than weeks. In our report above, it was more instructive to show gradual changes from graphene to fluorographene, which were easier to follow using the low $T$ fluorination. The use of higher $T$ is important for applications and, also, this proves that the FF state discussed above was final. Indeed, prolonged fluorination at 200 °C led to the same Raman, optical and transport characteristics as those achieved for very long exposures at 70 °C.

*Further experimental details:* The Raman studies were carried by using a Renishaw spectrometer with a green (514 nm) laser. For optical spectroscopy experiments we used a xenon lamp (250-1200nm) and Ocean Optics HR2000 spectrometer. Tecnai F30 TEM operated at 300 kV was employed for studies of the FG's structure. For micromechanical measurements, we used a MultiMode Nanoscope (*Veeco*) and, for electrical measurements, FG was transferred from a TEM grid back onto an oxidized Si wafer (step 5 in Figure 1). The standard microfabrication procedures[1,11] including electron beam lithography were then employed to make electrical contacts (step 6). More details are provided in the supporting online information.

**Note added at proof:** Our manuscript was considered by 8 referees in Nature series journals and eventually rejected by editors because they felt that a report[38] "scooped" the novelty. The latter appeared as advanced online publication 3 months after our initial submission. During this period, papers[12-14] were also published. As they were previously available online, we were able to discuss them in the Introduction. As for the latest report[38], it deals with graphene films grown on Cu and fluorinated by using $XeF_2$ at room *T*. The authors refer to their material as perfluorographane. It is a nonstoichiometric compound with a significant amount of fluorine adsorbed onto structural defects, which involves $C-F_2$, $C-F_3$ and other types of bonding as follows from the reported XPS data and discussed by the authors. This leaves many carbon sites within the graphene sheet itself non-fluorinated. The insulating and Raman properties of perfluorographane show the behavior similar to our samples of partially fluorinated graphene (approximately 30 hours in Figure 2), and we reach similar conclusions about these properties (see section 2.2 and 2.5 above and Supporting Information). The optical gap observed in our experiment is in agreement with calculations presented in ref.[38].

**Acknowledgements**
The work was supported by the Office of Naval Research, EPSRC (UK), the Körber Foundation, the Air Force Office of Scientific Research and the Royal Society.

**Supporting on-line material**

**#1. Fluorographene by exfoliation of graphite fluoride (GrF)**
Due to the layered nature of GrF, it is reasonable to try making its monolayers by mechanical exfoliation, the technique that proved successful for graphite and other layered materials [S1]. Recently, this approach has also been tried for GrF but no monolayers could be obtained [S2-S4]. In ref. [S2], GrF was reduced in solution to obtain *graphene* monolayers functionalized by hexane groups rather than fluorographene. In another report [S3], sonication of GrF allowed multilayer platelets (6-10nm thick) that were referred to as multilayer graphene fluoride [S3]. Similarly, ref. [S4] reports multilayer flakes of GrF, which contain "more than 10 monolayers". Raman, electrical and structural properties of these multilayer flakes were investigated and found to be close to those of bulk GrF [S3,S4]. In particular, as shown in the main text, the Raman spectra of GrF and its multilayers, reported in refs. [S3,S4], and even of GrF monolayers obtained in this work are characteristic to reduced fluorographene. Stoichiometric fluorographene exhibits a qualitatively different behavior.

To obtain monolayer FG that was studied in the present work we have used a different approach. This was partially because our efforts to obtain monolayers of GrF were also relatively unsuccessful and resulted only in small (micron-sized) and structurally damaged monolayers. Nonetheless, it could be helpful to briefly describe our efforts. Fig. S1A shows a photograph of our initial GrF material used for exfoliation. It is a white fine powder with a nominal composition $CF_{1.1\pm0.05}$ as measured by X-ray photoemission spectroscopy (XPS). The material was found to be extremely difficult to cleave down to individual layers. Only crystallites with a several nm thickness were in abundance, similar to reports [S3,S4]. Nonetheless, careful "hunting" in an optical microscope (on top of an oxidized Si wafer with 300 nm of $SiO_2$) allowed us to find a few examples of monolayers (see Fig. S1B). The monolayers give rise to little optical contrast and to locate them we concentrated on areas near thicker flakes. The contrast was a few % and mostly in blue (c.f. >10% for graphene). Consecutive AFM measurements showed that these regions were monolayers exhibiting a thickness of <2 nm above the substrate, similar to the AFM apparent thickness of single-layer graphene on $SiO_2$ [S1].

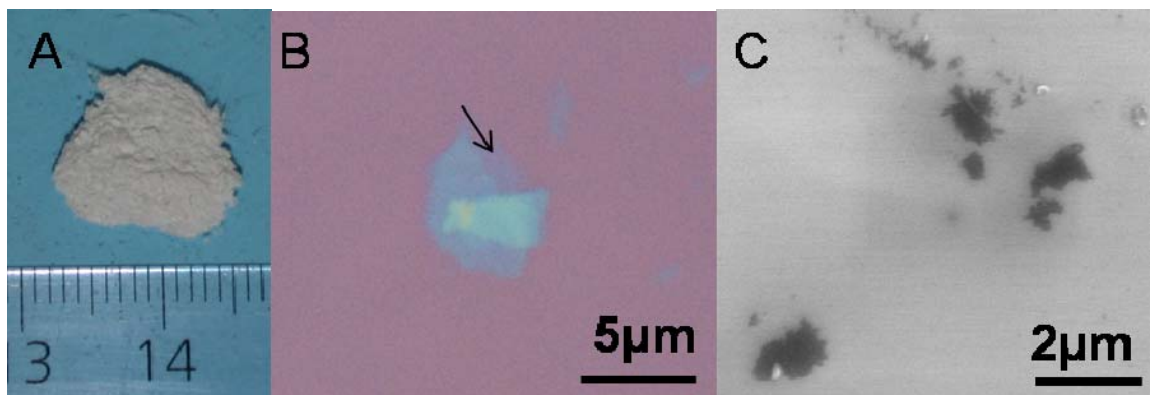

***Figure S1*** *(A) - Photograph of graphite fluoride. (B) – Thin layers of exfoliated GrF on an oxidized Si wafer. The arrow points at the region where a monolayer was identified by SEM and AFM. (C) Monolayer GrF gives rise to a strong contrast in SEM. The image was obtained at 5 keV using FEI Sirion.*

We found it easy to visualize GrF monolayers in SEM because their SEM contrast was even stronger than that of graphene (probably, due to GrF's high resistivity). Unfortunately, SEM provides no indication of the thickness of GrF (Fig. S1C). To this end, monolayers were identified as flakes visible in SEM but with a vanishingly little optical contrast. Our identification of cleaved GrF monolayers was confirmed retrospectively by using fluorographene obtained by XeF2 exposure, which exhibited the same optical, AFM and SEM characteristics. We attribute the difficulties of producing FG by mechanical cleavage to the small size of crystallites in the initial GrF material (small mesh powder is intentionally used in industry to enhance the speed of fluorination) and the fragility of monolayers because of the presence of a large number of structural defects. Note that no such difficulties were encountered to make monolayer from large (mm-sized) crystals of digraphite fluoride ($C_2F$) [S5]. This probably indicates that mechanical exfoliation can be successful if large high quality crystals of GrF are available.

#### #2. Fluorination of few-layer graphene
Figure S2 shows Raman spectra of bi- and few- layer graphene samples that were suspended on TEM grids and then fluorinated for several days using the same procedures as described in the main text. The intense D peak and the suppression of the 2D peak show that the fluorination reaction takes place even in multilayer samples. However, in comparison to monolayer graphene, the reaction is slow, which means the reactivity of the 2D material exposed from both sides is much higher than that of its 3D counterpart, in which fluorine has to diffuse between atomic planes. This model is in agreement with the observation of mesoscopic bubbles in such fluorinated samples, which can be attributed to molecular fluorine that is less reactive than F and trapped between atomic planes.

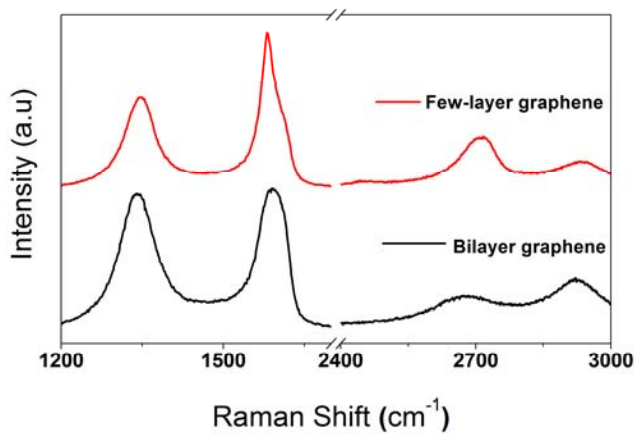

*Figure S2* Raman spectra of bi- and few- layer graphene after several days of exposure to atomic F obtained by decomposition of $XeF_2$ at 70 °C.

#### #3. Stability of fluorographene
This chapter provides additional information about stability of FG. Transport measurements discussed in the main text are sensitive to minute changes in the chemical composition but Raman spectroscopy provides a quick and non-destructive tool to evaluate more significant variations. The Raman analysis was carried out by using Renishaw spectrometer (wavelength of 514 nm) and graphene fluorinated in $XeF_2$ and then transferred onto an oxidized silicon wafer (step 5 in Fig. 1 of the main text). No changes could be detected in Raman signatures of FG after its exposure to various solvents (some are listed in the main text) and ambient air for many weeks. To induce changes in the FG composition, we annealed our samples at different $T$. Figure S3 shows Raman spectra for graphene fluorinated to different levels and then annealed in an argon-hydrogen (10%) mixture. One can clearly see from this figure that the stability of fluorinated monolayers strongly depends on the level of their fluorination. The D peak in weakly (1h) fluorinated graphene practically disappears after annealing at 250 °C, which indicates the reversibility of the initial chemical reaction (Fig. S3A), in agreement with results in ref. [S3]. For moderately fluorinated graphene (several h), the annealing at 250 °C led to a partial recovery with a strong D peak left afterwards. Attempts to anneal such samples at higher $T$ resulted in a further increase of the D peak, which could be attributed to structural defects formed when F was removed at high $T$. We have not seen any changes in the Raman spectra of FG at $T$ up to 250°C, except for the appearance of a luminescence background (with a broad peak centered at approximately 1.7 eV) (not shown). Prolonged annealing at 450°C led to the rise of the G and D peaks but the 2D peak did not recover (Fig. S3C). This probably indicates that the graphene scaffold becomes damaged with the loss of both C and F (similar Raman spectra were observed for FG after its long exposure to a 300 keV electron beam in TEM).

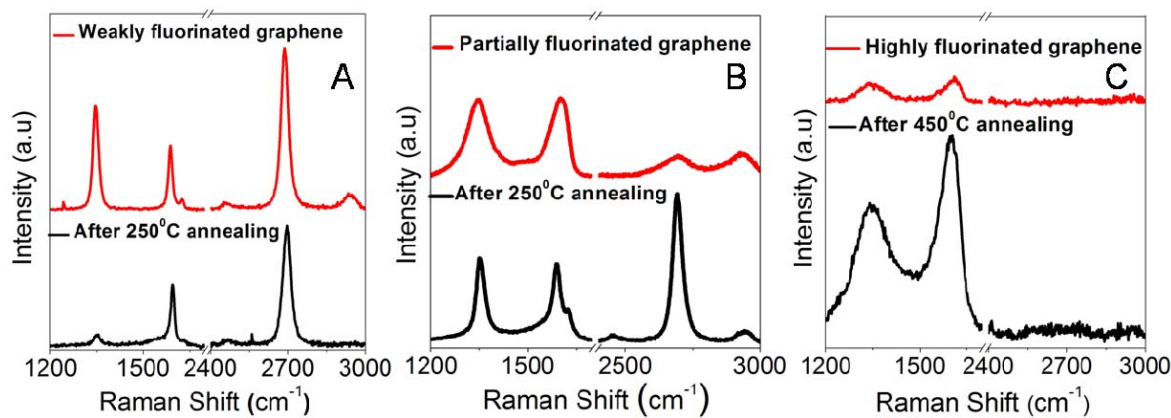

***Figure S3*** *Raman spectra of graphene fluorinated to various levels and then annealed at different T. (A,B,C) – Raman spectra for weakly, moderately and highly fluorinated graphene, respectively.*

#### #4. Chemical composition of fluorographene

The disappearance of all the Raman signatures for FG and the similarity between the spectra of GrF and graphene fluorinated for only 30 hours (see Fig. 2 of the main text) show that the fully fluorinated graphene should have a composition close to unity than GrF. The latter exhibits a fluorine-to-carbon ratio of ≈1.1. To find out more, we employed X-ray photoelectron spectroscopy and energy dispersive X-ray spectroscopy (EDX).

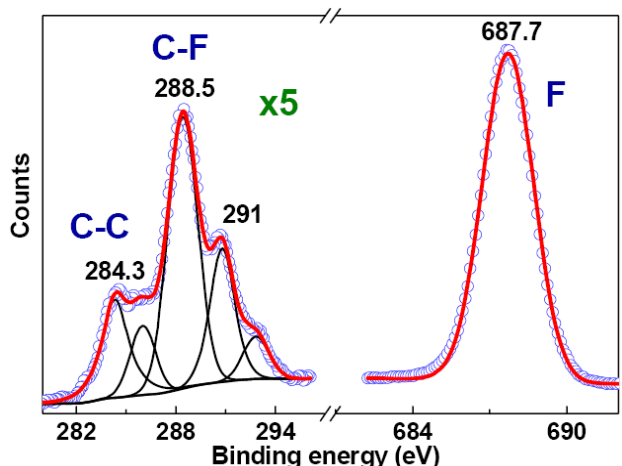

***Figure S4*** *XPS for graphene grown on SiC and fluorinated for two months in XeF$_2$ at 70 °C. Symbols are the measurements (carbon signal from SiC substrate is subtracted); solid curves the best fits.*

The XPS spectra of FG membranes revealed both F and C-F peaks indicating the extensive fluorination (F/C ratio ≥0.7) but the samples were too small for accurate composition analysis. Moreover, the supporting polymer (Quantifoil) scaffold was also fluorinated, which further obscured the XPS analysis. To circumvent the problem, we fluorinated large (cm-sized) areas of few-layer graphene grown on SiC. The fluorination process was monitored by Raman spectroscopy. It required two months of the exposure to XeF$_2$ to reach a saturated state with Raman spectra similar to those in the upper two curves in Fig. 2a (20 to 30h), depending on spot position. This was despite the fact that our pristine SiC samples exhibited a strong D peak, indicating many defects and grain boundaries, which should have enhanced diffusion of atomic F between graphene planes. Fig. S4 shows typical XPS spectra for such extensively fluorinated graphene on SiC, which - we emphasize - was still somewhat short of the FF state achieved for suspended graphene. One can see the pronounced F peak at 688 eV and the C-F peak at ≈289 eV. Their positions yield strong, covalent F bonding [S6]. The peak at ≈284 eV corresponds to C-C bonding, and the other peaks indicates the presence of different types of C-F bonding, including $CF_2$ (≈291 eV) and $CF_3$ (≈293eV). The spectrum shown in Fig. S4 yields an F/C ratio of ≈0.9. The ratio varied between 0.8 and 1.1 for different spots on our SiC samples, and the relative intensities of the C-F peaks also varied. F/C ratios larger than 1 are common for GrF and due to the presence of structural defects, which allow more C bonds to be terminated with fluorine ($CF_2$ and $CF_3$ bonding). The F/C ratios less than 1 can be attributed to the presence of partially fluorinated regions within an area of ≈100 μm in diameter that is probed by our XPS. This spatial inhomogeneity was also evidenced by Raman spectroscopy with numerous spots exhibiting spectra similar to the 20 hour curve in Fig. 2a. Despite the limitations caused by the incomplete fluorination of graphene on SiC, the XPS measurements provide the proof that graphene membranes, which allowed much quicker fluorination and exhibited weaker Raman signatures and no spatial inhomogeneity, contained more fluorine than graphene on SiC and, therefore, had a composition closer to stoichiometric than the latter.

We also investigated a local chemical composition of FG using EDX spectroscopy in a 300 kV TEM. Our sensitivity was not sufficient to accurately analyze a single layer of FG and, to circumvent the problem, we have prepared thicker samples by folding a FG monolayer approximately 10 times. Figure S5 shows its EDX spectrum and compares it with the corresponding spectrum of GrF ($CF_{1.1}$). Both spectra show the characteristic X-ray peaks for carbon and fluorine. What's more, FG has shown excellent spatial homogeneity of its EDX spectrum, whereas the spectra acquired from different positions on GrF samples exhibited some changes in the relative intensity of the fluorine peak. In particular, the peak was somewhat reduced when the electron beam was positioned close to GrF edges. The GrF spectrum in Fig. S5 is taken from a central part where the F peak is most intense. The similarity of the spectra for FG and GrF provides further proof that our FF graphene has a chemical composition very close to stoichiometric.

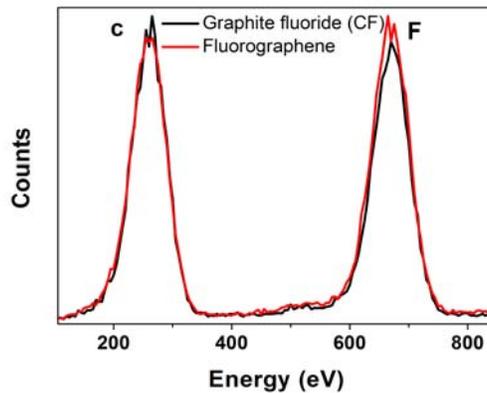

*Figure S5* EDX spectra for graphite fluoride and fluorographene.

**#5. Electron transport in weakly and moderately fluorinated graphene**
This regime has also been studied in a recent preprint [S4]. The electrical measurements of fluorinated graphene were carried out in the dc regime by using Keithley's 2410 SourceMeter and 2182A NanoVoltmeter. I-V characteristics and their *T* dependence were recorded for devices placed in a cryostat in a helium atmosphere. Similar devices (see insets of Figs. 5 and S6) but with no graphene sheet present showed a leakage current up to ~0.1 nA, if a high source-drain voltage of 10 V was applied. This is attributed to parasitic parallel resistances in the measurement circuit.

Figs. S6 and S7 extend the results reported in the main text by showing the electrical characteristics for graphene at smaller levels of fluorination. For weakly fluorinated graphene (1 hour at 70°C), our devices exhibited ρ in the MΩ range (Fig. S6) and their I-V characteristics remain linear for all measured *T* (>4K). We observed only a small increase in ρ with decreasing *T*. The devices exhibited strong donor doping (>$10^{13}$ cm$^{-2}$) and the electric field effect with a low mobility of 0.1 to 1 cm$^2$/Vs. This behavior can be explained by the presence of both fluorinated and pristine regions so that electron transport occurs mostly through the latter and involves lengthy percolation paths.

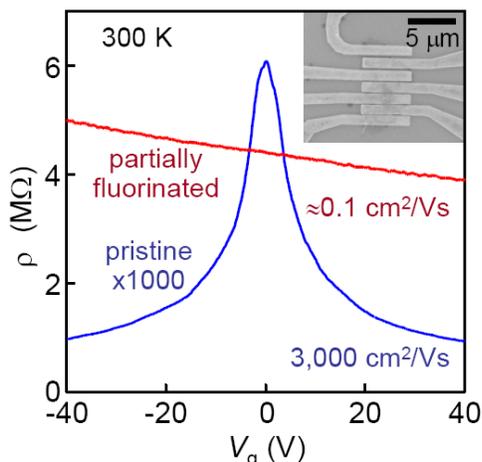

*Figure S6 Electron transport in a weakly (1 hour) fluorinated graphene. The upper (red) curve shows changes in its ρ as a function of back gate voltage. For comparison, the lower (blue) curve shows the electric field effect in pristine graphene. The inset shows an SEM image of one of our devices. The gaps between the Au contacts are between 50 and 500 nm, providing typical aspect ratios of 10 to 100.*

For the case of stronger fluorination, the device in Fig. S7 exhibited Raman spectra similar to the 9 hour curve in Fig. 2a. At room $T$, its I-V characteristics were linear and the resistance $R$ was well below 1 MΩ (corresponds to ρ ≈5 MΩ). At lower $T$, $R$ rapidly increased and I-V characteristics became strongly nonlinear below 50 K (Fig. S7). However, the Ohmic regime persisted at higher $T$ and low source-drain voltages. The inset in Fig. S7 plots the $T$ dependence of $R$ in this regime, which is well described by an Arrhenius-type behavior with an activation temperature of about 250 K. The value varied for different devices and fluorination levels. Note that the variable-range hopping dependence $\exp(\alpha/T^{1/3})$ can also describe these experimental data. It is clear that electron transport in such partially fluorinated samples involves hopping between impurity states but further studies would be required to find the exact transport mechanism (see ref. [S4]). Note that the transport behavior for our partially fluorinated graphene is rather similar to the one reported for thin platelets of reduced GrF with F/C ratio of ≈0.7 [S3] as well as for GO [S7].

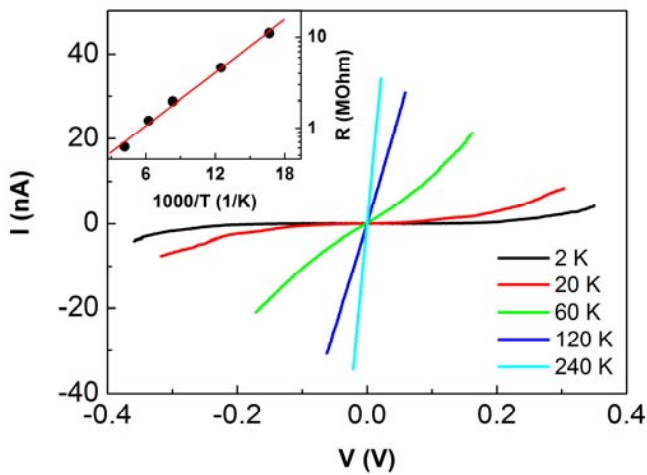

*Figure S7 I-V characteristics for a partially fluorinated graphene (nominally, 9h fluorination) at different T from 2 K to 240 K. Inset: resistance as a function of 1/T. The solid line is the best fit that corresponds to an activation temperature of about 250 K for this particular device.*

Finally, let us also mention that we tried to observe the electric field effect in the FF state but were unsuccessful. No onset of conductivity was found for gate voltages $V_g$ up to 100 V. This can be attributed by a large density ($>10^{13}$ cm$^{-2}$) of electronic traps (deep impurity states). The traps must be filled first, before mobile charge carriers appear, and this requires $V_g$ much higher than attainable for oxidized Si. The situation is rather similar to the case of many wide-gap semiconductors, in which no field effect could have been achieved so far. Also, one might recall the initial efforts to make silicon transistors, when considerable improvements took place to eliminate interface traps before the devices started exhibiting any field effect. Because of high thermal and chemical stability of FG, there might be ways to improve its electronic quality, which could in turn offer new venues for graphene electronics with the possibility of high on-off ratio FG transistors. However, we do not expect this development to be easy.

#### #6. AFM measurements of mechanical properties

This part of our work essentially repeats the studies of pristine graphene reported in ref. [S8]. To study the stiffness and breaking strength of FG we used a Veeco AFM (MultiMode Nanoscope) and tapping-mode doped silicon tips (Nanosensors PPP-NCHR). The tip radii were controlled by direct observation in SEM before and after the experiments. In total, 15 pristine and the same number of fully fluorinated membranes were investigated and then intentionally destroyed in the experiments. First the membranes (of diameter $D \approx 1.7\mu$m) were scanned in the tapping mode. Then the tip was positioned within $D/10$ from the centre of the membrane [S8]. The cantilever was then pushed into the sample until a threshold deflection was reached. The indentation $\delta$ of the membrane center was calculated from the difference between the cantilever deflection $d$ and the vertical tip movement $z$. The cantilever deflection was calibrated on a surface of silicon oxide which was assumed to be infinitely hard (that is, $d=z$). The load imparted on the membrane was calculated from the deflection of the cantilever and its effective spring constant $k$ ($\approx$40N/m for our cantilever) as $F=kd$. For small indentation depths (below the break point) no hysteresis between loading and unloading was observed and subsequent indentations were identical (Fig. S8). This shows that there was no slippage of the membranes relatively to the Quantifoil support during indentation. Breaking force $F_b$ was determined from the maximum bending $d_{max}$ of the cantilever before a membrane broke as $F_b=kd_{max}$ (Fig. S8). The breaking force was found to be in the range 250-800 nN for fluorinated and 900-1500 nN for pristine membranes (Fig. 6b of the main text). These spread in $F_b$ is similar to the one found for graphene membranes in ref. [S8].

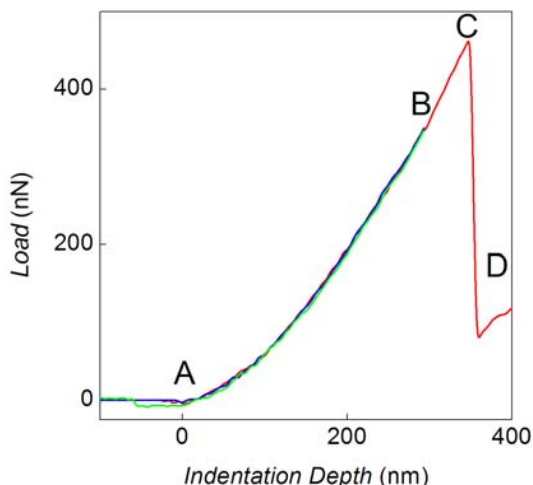

*Figure S8 Typical loading (blue&red) and unloading (green) curves for FG membranes. The first contact between the tip and the membrane happens at point A. Blue and green curves do not reach the maximum load and are non-hysteretic. If the load exceeded a certain limit (point C), the loading curve exhibited a sudden drop and was irreproducible in region D where the membrane was broken.*

To determine the Young modulus $E$, a model which involves two elastic membranes in series was employed. The reason for this is that the Quantifoil support has finite rigidity, so the indentation depth was partially spread between the support and the membrane. In the low load regime (load below 100 nN) we could fit the loading curves for both pristine and fluorinated graphene using a cubic dependence of the loading force on the membrane deflection [S8]. The deviations from the simple cubic behavior at higher loads can be attributed to elastic properties of the Quantifoil. By fitting the curves for pristine and fluorinated graphene with the same fitting parameters we have found the two-dimensional Young modulus for fluorographene of 100±30 N/m. The same parameters also yielded $E \approx 340$ N/m for our pristine graphene membranes, in agreement with ref. [S8].

#### #7. Opening the gap versus scattering in partially fluorinated graphene

To study the effect of disorder and gap opening we have performed numerical calculations of density of states (DOS) and optical conductivity $\sigma(\omega)$ for the model Hamiltonian

$$H = -t\sum_{\langle ij \rangle} c_i^+ c_j + v_d \sum_{i \in A} c_i^+ c_i - v_d \sum_{i \in B} c_i^+ c_i + \sum_i v_i c_i^+ c_i$$

at the honeycomb lattice. Here $\langle ij \rangle$ denotes the pairs of nearest neighbors, $2v_d$ is the energy gap between sublattices $A$ and $B$, $v_i$ is a random on-site potential uniformly distributed (independently on each site $i$) between $-v_r$ and $+v_r$ (further we will express all energies in units of the hopping parameter $t$). Calculations were performed for the crystallites 4096 by 4096 sites and 8192 by 8192 sites with periodic boundary conditions. To calculate DOS we used the method proposed in Ref. [S9]. The conductivity is defined by the Kubo formula [S10]

$$\sigma(\omega) = \lim_{\varepsilon \to +0} \frac{e^2}{(\omega + i\varepsilon)A} \left\{ -i[X,V] + \int_0^\infty dt\, e^{i(\omega+i\varepsilon)t} \langle [V,V(t)] \rangle \right\},$$

where $A$ is the sample area, $X = \sum_i X_i c_i^+ c_i$ is the coordinate operator ($X_i$ is the $x$-coordinate of site $i$), $V = \frac{i}{\hbar}[H,X]$ is the velocity operator.

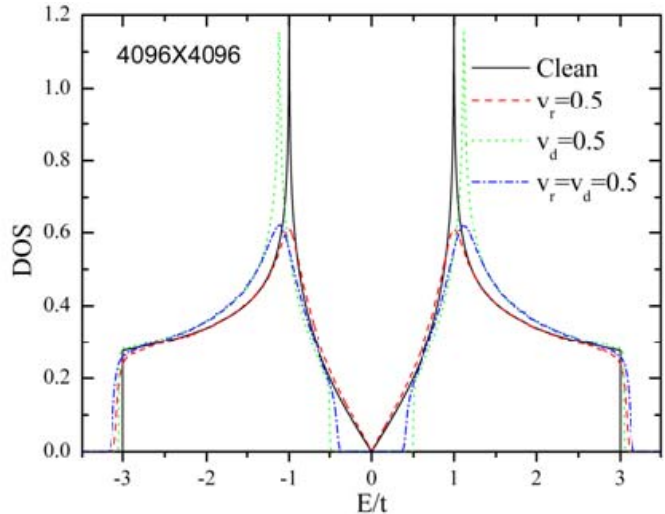

*Figure S9* Density of states. Solid black curve – pristine graphene without an energy gap ($v_r = v_d = 0$); Green dotted line – gapped clean graphene; Red dashed line – disordered graphene without a gap; Blue dashed-dotted line – disordered and gapped case. All energy parameters are in units of t and the DOS is in units of 1/t.

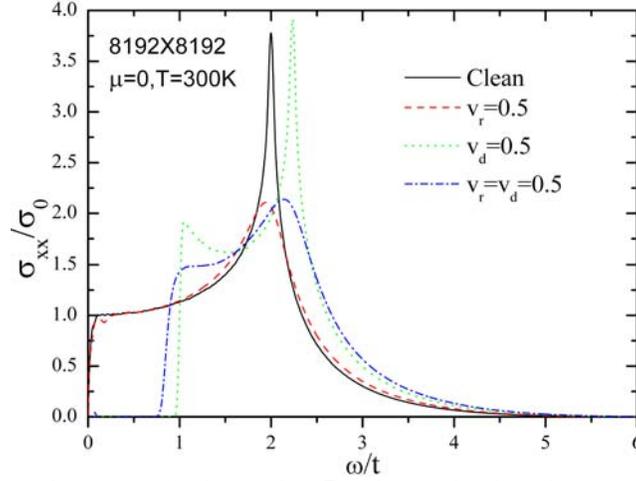

*Figure S10* Real part of conductivity (without the Drude peak) for the case of graphene without a gap ($v_r = v_d = 0$, solid black line); clean gapped graphene (green dotted line); disordered graphene (red dashed line); and the disordered gapped case (blue dashed-dotted line). The calculations were carried out for chemical potential μ = 0 and temperature T = 300K. $\sigma_0 = \frac{\pi}{2}\frac{e^2}{h}$ is the low-frequency optical conductivity of clean honeycomb lattice [S12].

For noninteracting fermions, after some manipulations, the Kubo formula can be transformed to the expression [S11]

$$\sigma(\omega) = D\left[\delta(\omega) + \frac{i}{\pi\omega}\right] + \lim_{\varepsilon \to +0} \frac{e^2 g(\omega)}{(\omega + i\varepsilon)A},$$

$$g(\omega) = \frac{1 - e^{-\beta\hbar\omega}}{\hbar\omega} \int_0^\infty dt\, e^{i(\omega+i\varepsilon)t} 2i\,\mathrm{Im}\langle V[1 - f(H)]V(t)f(H)\rangle \quad (S1)$$

Equation (S1) was used in our computations. Here $D$ is the Drude weight (we do not present the expression and results for it, because we are interested here only in light adsorption at finite ω), $\beta = 1/T$ is the inverse temperature and $f(H) = \frac{1}{\exp[\beta(H - \mu)] + 1}$ is the Fermi-Dirac function of the operator $H$. The angular brackets in Eq. (S1) means the trace over the whole Hilbert space of the system divided by the number of states. Similar to Ref. [S9] we replace this by an average over a single function $|\phi\rangle = \sum_i a_i |i\rangle$ where $a_i$ are random complex numbers. The time evolution operator and Fermi-Dirac operator were represented as the Chebyshev polynomial expansions. The results are shown in Figs. S9 and S10 (for the conductivity, we present only the real part determining the light adsorption, without the Drude peak at ω = 0).